\documentclass[12pt]{article}
\usepackage{a4,epsfig}

\begin{document}

\begin{titlepage}

\pagenumbering{arabic}
\vspace*{-1.5cm}
\begin{tabular*}{15.cm}{lc@{\extracolsep{\fill}}r}
&
\end{tabular*}
\vspace*{2.cm}
\begin{center}
\Large
{\bf \boldmath
    Influence of the bin-to-bin and inside bin correlations
    on measured quantities
%
%
%
} \\
\vspace*{1.4cm}
\normalsize

 {\bf V.~Perevoztchikov}$^{1}$,
 {\bf M.~Tabidze}$^2$
   {\bf~and~A.~Tomaradze}$^{3}$\\
 {\footnotesize $^1$ Brookhaven National Laboratory, Upton, USA}\\
 {\footnotesize $^2$ Institute for High Energy Physics of Tbilisi State University, Georgia}\\
{\footnotesize $^3$ Northwestern University, Evanston, USA. e--mail Amiran@fnal.gov}

\vspace*{0.6cm}
\normalsize {

}
\end{center}
\vspace{\fill}
\begin{center}
\end{center}
\vspace*{1.cm}
\begin{abstract}
\noindent

A new method for measuring the quantities influenced by
bin-to-bin and inside bin correlations is presented.
This method is essential for large multiplicity and/or
high density of particles in phase space.
The method was applied to  the two particle correlation
functions of $e^+e^-\rightarrow$W$^+$W$^-$ events.

\noindent
\end{abstract}
\vspace{\fill}
%

%
%
\end{titlepage}




\setcounter{page}{1}



\section{Introduction}

 The usual calculations of statistical errors for entries in
histograms and the use of these errors in the fitting procedure
can  bias the measurements if there are several entries
from the same event. Traditionally, in the past this problem was  ignored.
The effect is small for low multiplicity events. However, for LEP and
especially for the future LHC, RHIC, etc... experiments this effect is
not small at all.
These entries are correlated creating bin-to-bin and
inside bin correlations.  Neglecting  these correlations,
as we will show below, leads to a remarkable  underestimation of
the errors in the measured quantities and less precise estimation of the
quantities themselves.

An effective approach has been proposed in~\cite{deangelis}.
The exact method decribed here includes all possible correlation effects.

The method was applied to the two-particle
correlation functions of $e^+e^-\rightarrow$W$^+$W$^-$ events.

\section{The Method}

In this section a new method  is presented for construction of  the
covariance matrix between the bins of the histogram for an unbiased
measurement of the fitted quantities.

 The presence of bin-to-bin correlations in two particle
distributions is unavoidable. If there are
$N$ positive tracks in the events, each of them has $(N-1)$ entries
in the two-particle density $P$, contributing to different bins
of the histogram. Also, due to the finite bin width, the same track can
also enter several times in the same bin, which is a source of
inside bin correlations.

The method is based on classical statistics.
Let us consider the $i$-th event from the set of $n$ events and the
two-particle density $P$ which is presented in the histogram $h^i$
with $N_{p}$ bins.

 The histogram  $ H = \sum\nolimits_{i=1}^{n}h^{i} $ and values
\begin{eqnarray*}
  c_{jk}=\sum\nolimits_{i=1}^{n}
         (h^i_j-H_{j}/n)(h^i_k-H_{k}/n)(1+1/n) \
\end{eqnarray*}
were calculated  event by event.  Here $j$ and $k$ are the bin numbers
for the  histograms.
We do not know the correlations and errors for one event. But we know
that the different events are uncorrelated. Let us consider bin values
of the histogram made for one event as an random vector with unknown
distribution. We have an uncorrelated ensemble of these vectors and
hence we can estimate the covariance matrix statistically.
It is important to note, that this algorithm computes both
the ``technical'' and the physical correlations.
For all events we have the resulting histogram H for the
two-particle density $P$ and $ V_{jk}=c_{jk}\cdot n/(n-1) $  covariance
matrix for this histogram.
Note, that the expresion for unbiased estimation of sample
covariance was used for the calculation of covariance between
two values\cite{Korn}.
The diagonal terms $V_{jj}$ of the covariance matrix are assumed
to be the estimate for the squares of the error $\sigma_j$ of the
 $j$-th of histogram $H$.

\section{The Test of the Method Using the Simulated Events}

The precise measurement of the correlations between particles became
important for  $e^+e^-\rightarrow$W$^+$W$^-$ events due
to possible large impact of these correlations on the measured
W mass \cite{lep2be}. Thus, we applied
the method to the correlation functions in fully hadronic
and semileptonic WW events.

The correlation function $R$ is used to
study the enhanced probability for emission of particles.
For pairs of particles, it is defined as
\begin{equation}
R(p_{1},p_{2}) = \frac{P(p_{1},p_{2})}{P_{0}(p_{1},p_{2})} \, ,
\end{equation}
where $P(p_{1},p_{2})$ is the two-particle probability density,
 $p_{i}$ is the four-momentum of
particle $i$, and $P_{0}(p_{1},p_{2})$ is a
reference two-particle distribution which,
ideally, resembles $P(p_{1},p_{2})$ in all respects, apart from the lack
of Bose-Einstein symmetrization.
The effect is usually described in
terms of the variable $Q$, defined by
$Q^{2}=-(p_1-p_2)^2=M^{2}(\pi \pi)-4m^{2}_{\pi}$,
where $M$ is the invariant mass of the two pions.
The correlation function can then be written as
\begin{equation}
R(Q) = \frac{P(Q)}{P_{0}(Q)},
\end{equation}
which is frequently parametrised by the function
\begin{equation}\label{beq}
R(Q)=  1 + \lambda e^{-r^2Q^2}  \, .
\end{equation}
In the above equation, in the hypothesis of a spherically
symmetric pion source, the parameter $r$ gives
the RMS radius of the source and $\lambda$ is
the strength of the correlation between the pions.

The method described in section 2 was tested on the JETSET
simulation \cite{PYTHIA}.
We choose the fully hadronic and semileptonic
decay of WW pairs with  Bose-Einstein correlations included using
the LUBOEI code.  The value of  $\lambda=0.85$ and $r=0.5$~fm was used.
In case of fully hadronic channel the Bose-Einstein correlations
were switched on for all pions(full Bose-Einstein correlations).

The correlation
matrices $\rho_{jk}=V_{jk}/(\sigma_j\sigma_k)$ for like-sign pairs of
WW semileptonic channel (refered as (${2q}$) mode) and for fully hadronic
channel (referred as (${4q}$) mode), computed using 100 000 simulated
events for each sample, are shown in Fig.~1 and Fig.~2.
The correlations between bins for the WW fully hadronic
channel(Fig.~2) are larger than in the mixed hadronic and leptonic
channel(Fig.~1). Thus, the bin-to-bin correlations
are increased with multiplicity, as expected. Notice that the
correlations are nearly
independent of Q, which shows why the effective approach of~\cite{deangelis}
is a good approximation in this case.

For the future analysis the 500 samples of
3000 events each for (${4q}$) channel and 500 samples of 1500
events each for (${2q}$) channel were simulated.
For each of these samples  a histogram of the correlation function $R(Q)$ was
built, using the 25 bins of 100 MeV from 0 to 2.5 GeV.
The simulated $R$ distributions were
normalized to unity in the region $Q >$ 0.8 GeV/$c^2$.
We  performed a $\chi^2$ fit to the $R(Q)$ to the form (3)
for each of 500 samles.
%

The average values of $\lambda$ and $r$ from these ``naive'' fits were:
\begin{eqnarray}
\lambda_{2q}&=&0.333 \pm 0.029 , \\
r_{2q}&=&0.562 \pm 0.033   \mbox{ fm} \
\end{eqnarray}
for (${2q}$) events and
\begin{eqnarray}
\lambda_{4q}&=&0.416 \pm 0.013 ,  \\
r_{4q}&=&0.561 \pm 0.012   \mbox{ fm} \
\end{eqnarray}
for (${4q}$) events.  The statistical errors correspond to
the average of the errors for 500 samples.
The ``pull'' of the fitted values of $\lambda$
and $r$ for the simulated (${4q}$)
events are shown in Fig.~3.
A Gaussian fits gave that the errors in parameres
$\lambda$ and $r$ are underestimated by a factor 1.20 $\pm$ 0.05
and 1.30  $\pm$ 0.05
for the (${2q}$) events and by a factor
1.42 $\pm$ 0.06  and 1.53 $\pm$ 0.06  for the (${4q}$) events.

The average values of $\lambda$ and $r$ from the 500 fits using the
inverted $V_{jk}$ matrix (calculated for each of 500 samples) were:
\begin{eqnarray}
\lambda_{2q}&=&0.332 \pm 0.034 , \\
r_{2q}&=&0.556 \pm 0.040   \mbox{ fm} \
\end{eqnarray}
for (${2q}$) events and
\begin{eqnarray}
\lambda_{4q}&=&0.403 \pm 0.017 ,  \\
r_{4q}&=&0.565 \pm 0.016   \mbox{ fm} \
\end{eqnarray}
for (${4q}$) events.
The ``pull'' of the fitted values of $\lambda$
and $r$ for the simulated (${4q}$)
events are shown in Fig.~4.
A Gaussian fits gave
$\sigma_{(\lambda-<\lambda>)}/\sigma_{\lambda}$=1.06 $\pm$ 0.05 and
$\sigma_{(r-<r>)}/\sigma_{r}$=1.02 $\pm$ 0.05 for (2q) channel, and
$\sigma_{(\lambda-<\lambda>)}/\sigma_{\lambda}$=0.96 $\pm$ 0.05 and
$\sigma_{(r-<r>)}/\sigma_{r}$=1.08 $\pm$ 0.05 for (4q) channel.
The above values are  in a good agreement with unity and
thus the errors are correctly estimated.

\section{Summary}

A model independent method for measuring  the quantities
influenced by bin-to-bin and inside bin correlations is described.
A package, as an addition to HBOOK, was written to support this
new functionality. The method was tested using the simulated WW events.

\newpage
\setlength{\unitlength}{0.7mm}


\begin{figure}[hbt]
\begin{center}
\mbox{\epsfxsize16.0cm\epsfysize19.0cm\epsffile{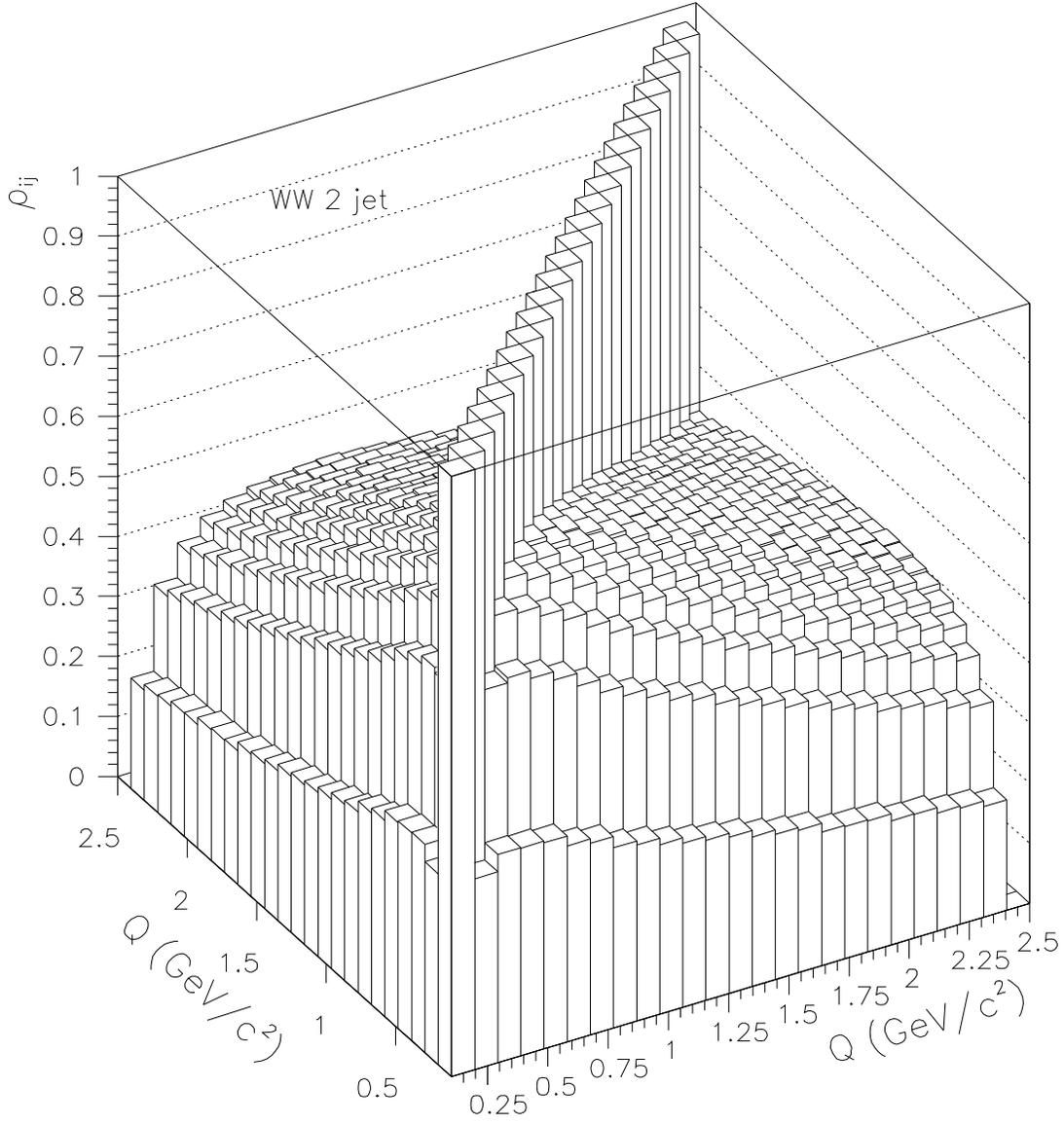}}
\end{center}
\caption{
The correlation matrix for like-sign pairs obtained using
the simulated  WW $({2q})$ events.
}
\end{figure}

\begin{figure}[hbt]
\begin{center}
\mbox{\epsfxsize16.0cm\epsfysize19.0cm\epsffile{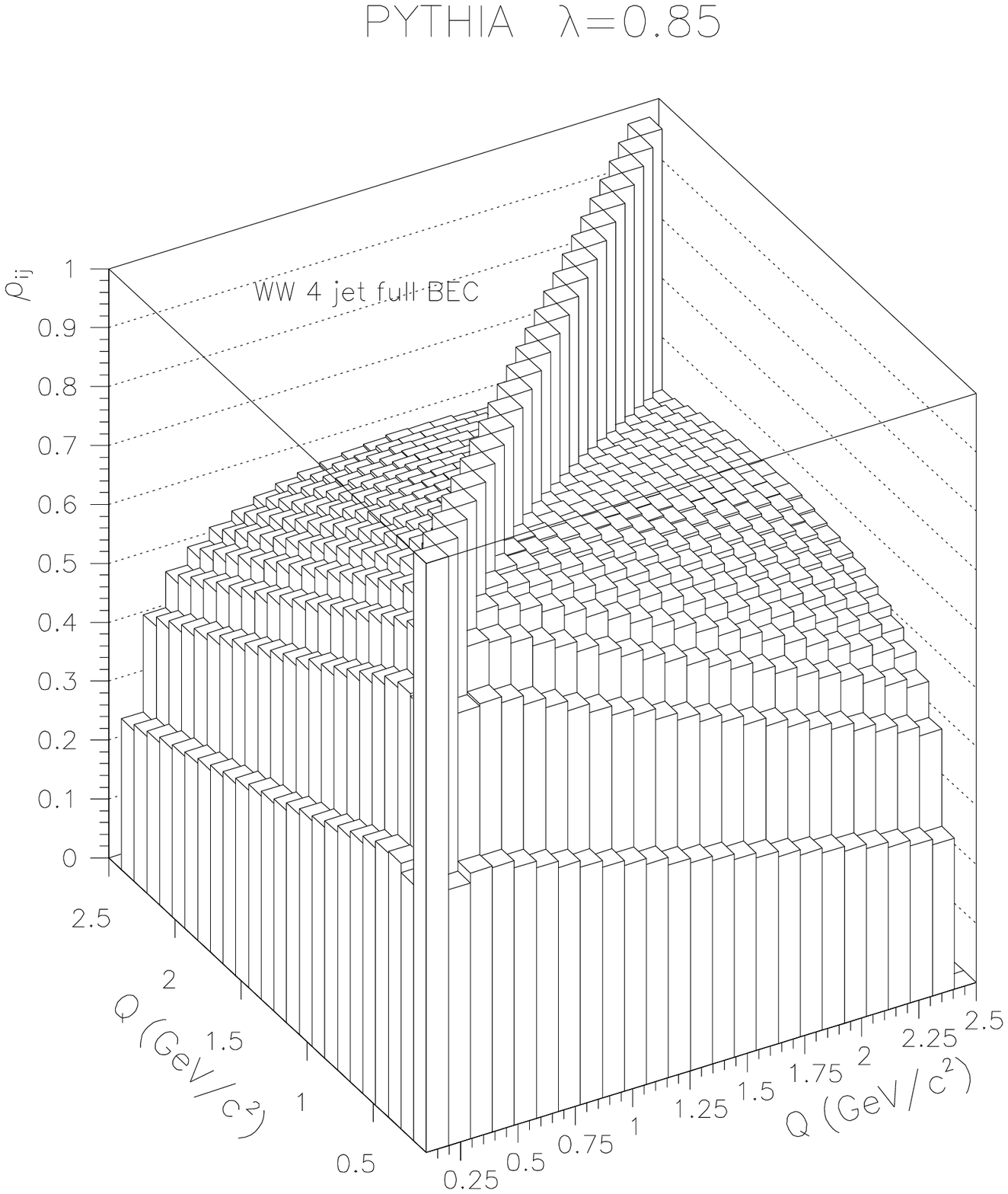}}
\end{center}
\caption{
The correlation matrix for like-sign pairs obtained using
the simulated  WW $({4q})$ events.
}
\end{figure}

\begin{figure}[hbt]
\begin{center}
\mbox{\epsfxsize16.0cm\epsfysize19.0cm\epsffile{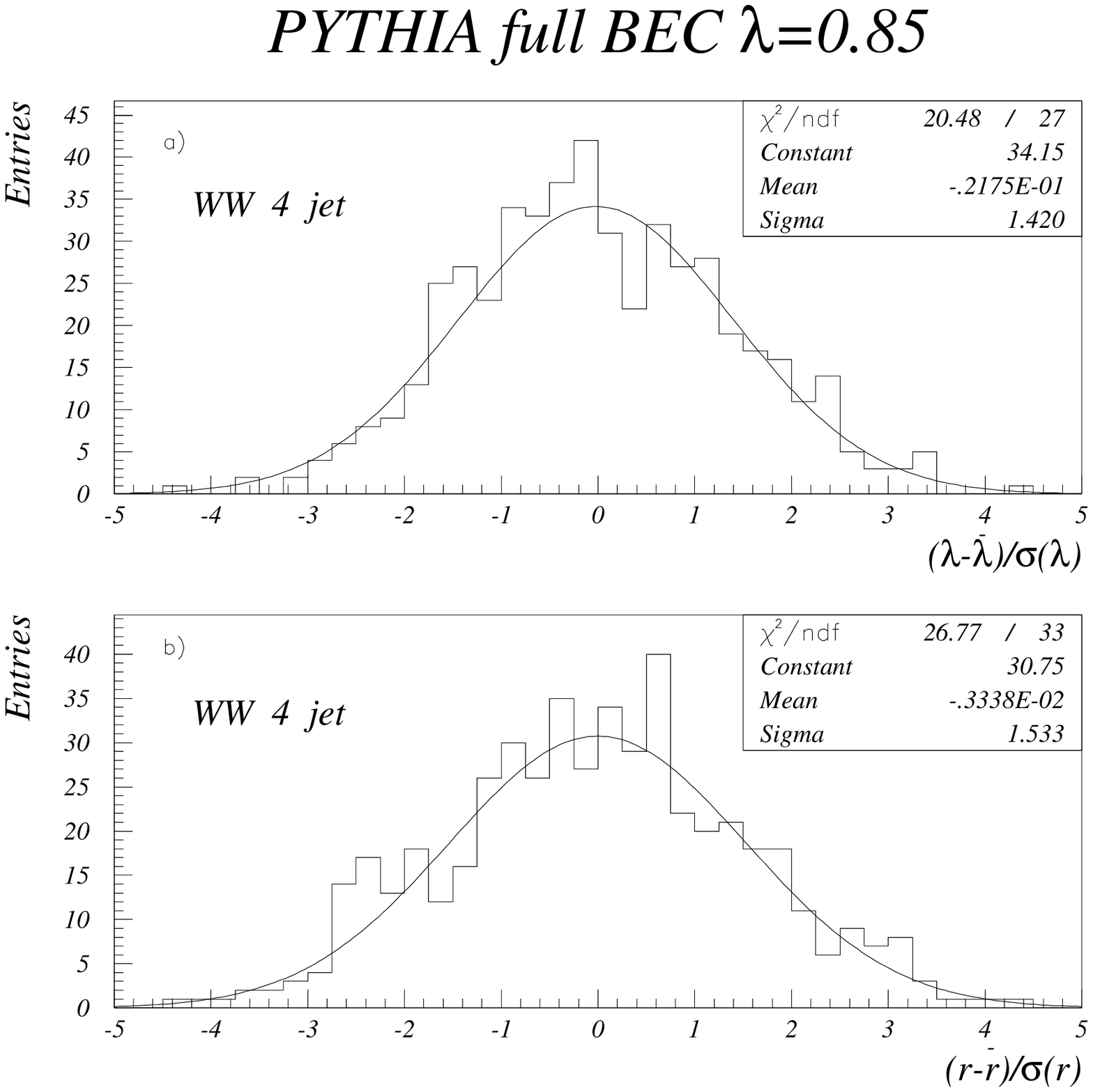}}
\end{center}
\caption{
 (a) Pull function for fitted parameter $\lambda$ using a binned uncorelated
least squares fit in the simulated samples. A gausian fit is supperimposed as
a solid line. b) Same as (a) but for the parameter $r$.
}
\end{figure}

\begin{figure}[hbt]
\begin{center}
\mbox{\epsfxsize16.0cm\epsfysize19.0cm\epsffile{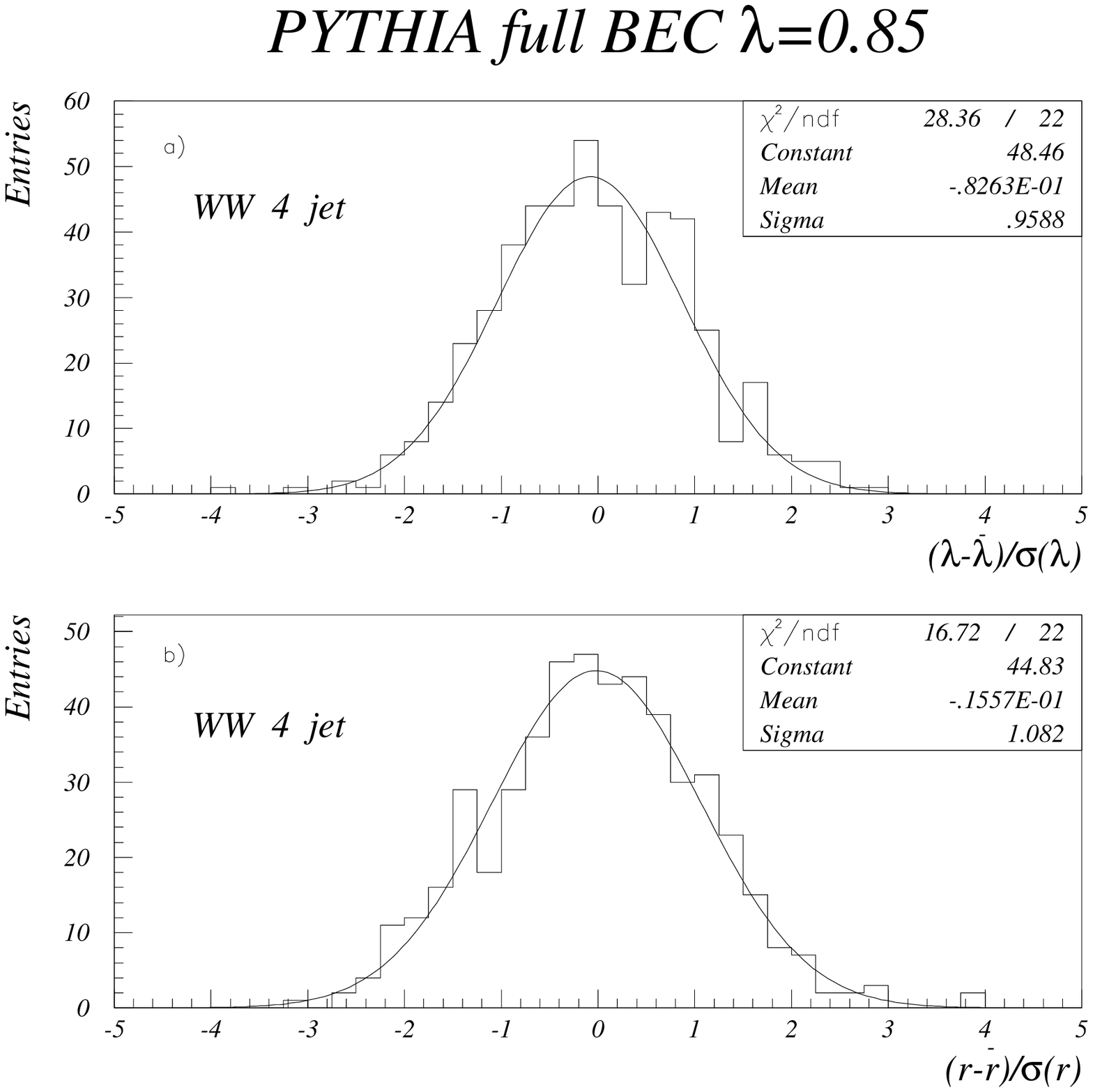}}
\end{center}
\caption{
 (a) Pull function for fitted parameter $\lambda$ in the simulated samples
using  covariance matrix technique. A gausian fit is supperimposed as
a solid line. b) Same as (a) but for the parameter $r$.
}
\end{figure}


\vspace{1cm}
\begin{thebibliography}{99}
\itemsep 0pt
\parsep 0pt

\bibitem{deangelis} A.~De Angelis and L. Vitale,
         Nucl. Instr. Methods {\bf A423} (1999) 446.


\bibitem{Korn} A.~Korn and M.~Korn.  "Mathematical handbook for
scientist and engineers", New York, 1968.

\bibitem{lep2be}
 L.\ L\"onnblad and T.\ Sj\"ostrand,
 Eur. Phys. J. {\bf C2} (1998) 165.
%

\bibitem{PYTHIA} T.\ Sj\"ostrand, Comp.\ Phys.\ Comm.\ {\bf 82} (1994) 74;
T.\ Sj\"ostrand et al., Comp.\ Phys.\ Comm.\ 135 (2001) 238;
for more details see T.\ Sj\"ostrand,  L.\ L\"onnblad and S. Mrenna,
{\it PHYTIA 6.2 Physics and Manual,} hep--ph/0108264.


\end{thebibliography}
\end{document}